\documentstyle[11pt,epsf]{article}
\newcommand{\Kbar}{\not{\!K}}

\newcommand{\Pbar}{\not{\!P}}

\newcommand{\be}{\begin{equation}}
\newcommand{\ee}{\end{equation}}
\newcommand{\ba}{\begin{eqnarray}}
\newcommand{\ea}{\end{eqnarray}}

\newcommand{\nsigma}{\mbox{\boldmath $\sigma$}}
\newcommand{\ngamma}{\mbox{\boldmath $\gamma$}}

\newcommand{\ntau}{\mbox{\boldmath $\tau$}}
\newcommand{\npi}{\mbox{\boldmath $\pi$}}
\newcommand{\nrho}{\mbox{\boldmath $\rho$}}

\newcommand{\nk}{{\bf      k}}
\newcommand{\np}{{\bf      p}}       

\newcommand{\nr}{{\bf      r}}

\newcommand{\columnmatrix}[2]{\left[
                               \begin{array}{cc}
                                  \displaystyle #1 \\[1ex]
                                  \displaystyle #2
                               \end{array}
                            \right]}
\begin{document}
\begin{titlepage}
\mbox{} 
\vspace*{2.5\fill} 
{\Large\bf 
\begin{center}
%
Momentum distribution of relativistic nuclei with Hartree-Fock mesonic
correlations
%
\end{center}
} 
\vspace{1\fill} 
\begin{center}
{\large 
J.E. Amaro$    ^{1}$, 
M.B. Barbaro$  ^{2,3}$, 
J.A. Caballero$^{3}$, 
T.W. Donnelly$ ^{4}$ and 
A. Molinari$   ^{2}$
}
\end{center}
\begin{small}
\begin{center}
$^{1}${\sl 
Departamento de F\'\i sica Moderna,
Universidad de Granada, 
E-18071 Granada, SPAIN 
}\\[2mm]
$^{2}${\sl 
Dipartimento di Fisica Teorica,
Universit\`a di Torino and
INFN, Sezione di Torino \\
Via P. Giuria 1, 10125 Torino, ITALY 
}\\[2mm]
$^{3}${\sl 
Departamento de F\'\i sica At\'omica, Molecular y Nuclear \\ 
Universidad de Sevilla, Apdo. 1065, E-41080 Sevilla, SPAIN 
}\\[2mm]
$^{4}${\sl 
Center for Theoretical Physics, Laboratory for Nuclear Science 
and Department of Physics\\
Massachusetts Institute of Technology,
Cambridge, MA 02139, USA 
}
\end{center}
\end{small}

\kern 1. cm \hrule \kern 3mm 

\begin{small}
\noindent
{\bf Abstract} 
\vspace{3mm} 

The impact of Hartree-Fock correlations on the nuclear momentum
distribution is studied in a fully relativistic one boson exchange
model. Hartree-Fock equations are exactly solved to first order
in the coupling constants. The renormalization of the Dirac spinors in
the medium is shown to affect the momentum distribution, as opposed to
what happens in the non-relativistic case. The unitarity of the model 
is shown to be preserved by the present renormalization procedure.

\kern 2mm 

\noindent
{\em PACS numbers:}\  21.60.Jz,  21.65.+f, 24.10.Jv, 11.10.Gh

\noindent
{\em Keywords:}\ Relativistic Hartree Fock. 
Renormalization in nuclear matter.
Mesonic correlations.  Relativistic Fermi Gas.

\end{small}
\kern 2mm \hrule \kern 1cm
\noindent MIT/CTP\#3274 
\end{titlepage}

It is well known that in a non-relativistic framework the momentum
distribution of nuclear matter is not affected by the Hartree-Fock
(HF) field. This arises because, due to general invariance principles
\cite{Fet71}, the non-relativistic self-energy cannot depend on spin
in an infinite system: hence the single-nucleon wave functions are not
modified and only the energy-momentum relation is affected by the
medium.  Of course correlations in the nuclear wave function beyond
the mean field approximation are very important already at the
non-relativistic level~\cite{Mut00,Fan84,Dic92}.  Due to such correlations,
the momentum distribution is reduced for momenta below $k_F$ and the
states above $k_F$ acquire small but finite occupation probabilities.

Although the momentum distribution is not an observable, it is
also true that over the years electron scattering reactions have frequently been expressed in terms of momentum densities. 
In recent work~\cite{Ama01,Ama02}
we have evaluated the impact of mesonic
correlations and meson-exchange currents (MEC) on the electroweak response functions within a fully
relativistic, gauge invariant model. We have shown that the
consistency of the theory necessarily implies the inclusion in the calculation
of Hartree-Fock self-energy insertions. In order to deal properly with the 
divergencies associated with these diagrams, not only the
energy but also the nucleon wave functions must be renormalized 
by the medium. As a consequence, in a relativistic HF framework, 
the momentum distribution is also modified for $k<k_F$, since now the
Dirac four-spinors describing the nucleons display new features
and the self-energy becomes spin-dependent.  The aim of this letter is 
to quantify this genuine 
relativistic effect in a one boson-exchange model for the NN
interaction, while the
corresponding observable consequences on the response functions were
analyzed in depth in~\cite{Ama01,Ama02}.

An unambiguous treatment of relativistic Hartree-Fock does not exist
in the literature, since the presence of the Dirac sea requires (at least) the specification of a prescription to take it into account~\cite{Ser86}. 
The approach we use is equivalent to that used in~\cite{Ser86,Cel86}, where
the nucleon proper self-energy is calculated in terms of
positive-energy spinors only. 
This approximation is valid in the
first iteration of a fully self-consistent calculation to which we
confine ourselves in this work. 
This procedure was shown in~\cite{Ser86} to reproduce the
non relativistic HF equations in the limit $M \rightarrow \infty$, and
it reduces the relativistic Hartree approximation to the
Mean Field Theory.

The proton momentum distribution of nuclear matter in the
independent particle approximation is
\be
n(\np) = \sum_{\nk,s} 
\psi_{\nk,s}^\dagger(\np)\psi_{\nk,s}(\np) \theta(k_F-k)\, \ ,
\label{np}
\ee
where $k_F$ is the Fermi momentum.
Since we are focusing on symmetric nuclear matter the neutron and
proton momentum distributions are equal.
For a free relativistic Fermi gas, in momentum space 
the wave function describing a nucleon with momentum $\nk$ and spin $s$
is given by
\be
\psi_{\nk,s}(\np) = \int_V d\nr\, \psi_{\nk,s}(\nr) e^{-i\np\cdot\nr} 
= \sqrt{\frac{m}{V E_\nk}}\, u_s(\nk,m) \int_V d\nr e^{i(\nk-\np)\cdot\nr} 
= \sqrt{\frac{V m}{E_\nk}}\, u_s(\np,m) \delta_{\nk,\np}
\label{psifree}
\ee
where  $V$ is the volume enclosing the system, 
$u_s(\nk,m)$ is the free Dirac spinor  ($\Kbar u_s =m u_s$) and
$E_\nk=\sqrt{\nk^2+m^2}$ is the free energy of the nucleon. 
We use the Bjorken and Drell~\cite{Bjo65} conventions for the spinor
normalization, $\bar u u=1$. Therefore
the wave function in coordinate space is normalized
to one: $\int_V d\nr\psi_{\nk,s}^\dagger(\nr)\psi_{\nk,s}(\nr)=1$.

The wave function of Eq.~(\ref{psifree}), inserted in Eq.~(\ref{np}), 
yields the well-known result:
\be
n(\np) 
= V \sum_{\nk,s}\frac{m}{E_\nk} 
u_{s}^\dagger(\np,m) u_{s}(\np,m)\theta(k_F-p)\delta_{\np,\nk}
= 2 V \theta(k_F-p)\ .
\label{npfree}
\ee

In an interacting system, in the relativistic HF approximation, 
the above distribution is modified, since the 
single-particle wave functions are renormalized by the interaction
with the other nucleons in the medium. 
In this case the
Dirac equation in the nuclear medium is given by
\begin{equation}\label{new_spinors}
[\Pbar-m-\Sigma(P)]\widetilde \phi_s(\np)=0\ ,
\end{equation}
where $\widetilde\phi_s(\np)$ is the renormalized spinor and 
$\Sigma(P)$ is the self-energy of a nucleon in nuclear matter.
According to general symmetry properties $\Sigma(P)$ can be written in
the form~\cite{Cel86,Ana81}:
\begin{equation}
\label{spin}
\Sigma(P) = mA(P)+B(P)\gamma_0 p_0 -C(P)\ngamma\cdot\np \ .
\end{equation}

Using the above decomposition the Dirac equation (\ref{new_spinors})
can be recast as 
\begin{equation}
\label{Dirac}
\left[1-C(P)\right]
\left[ \gamma_0 f_0(P)-\ngamma\cdot\np-
\widetilde{m}(P)\right]\widetilde \phi_s(\np)=0\ ,
\end{equation}
where the functions
\begin{eqnarray}
\label{f0}
f_0(P) &=& \frac{1-B(P)}{1-C(P)}\ p_0 \label{f_0}\\
\widetilde{m}(P) &=& \frac{1+A(P)}{1-C(P)}\ m  \label{m-tilde}
\end{eqnarray}
have been introduced.

Equation~(\ref{Dirac}) has the same structure as the free Dirac equation;
hence for the positive-energy eigenvalue one has
\begin{equation}
f^2_0(P)= \np^2+\widetilde{m}^2(P)\ ,
\end{equation}
which implicitly yields, using Eq.~(\ref{f0}),
the new dispersion relation for the renormalized energy 
$p_0=\epsilon(\np)$ of the nucleon in the nuclear medium:
\begin{equation}
\label{dispersion}
p_0 = \frac{1-C(P)}{1-B(P)}\sqrt{\np^2+\widetilde{m}^2(P)}\ .
\label{p0eq}
\end{equation}

The corresponding positive-energy spinor 
reads (see Refs.~\cite{Ama01,Ama02} for details)
\begin{equation}
\widetilde\phi_s(\np)\equiv\widetilde u_s(\np,\widetilde m(\np)) = 
\sqrt{Z_2(\np)}
\left(
\frac{\widetilde{E}(\np)+\widetilde{m}(\np)}{2\widetilde{m}(\np)}
\right)^{1/2}
\columnmatrix{\chi_s}{
\frac{\nsigma\cdot\np}{\widetilde{E}(\np)+\widetilde{m}(\np)}\chi_s}
=
\sqrt{Z_2(\np)}u_s(\np,\widetilde{m}(\np))\ ,
\label{phispin}
\end{equation}
where the function $\widetilde{m}(\np)$ of the three-momentum $\np$
is obtained from the Dirac mass in 
Eq.~(\ref{m-tilde}) by setting $p_0=\epsilon(\np)$:
\be
\widetilde{m}(\np)
\equiv
\widetilde{m}(\epsilon(\np),\np) 
\label{Dirac-mass}
\ee
and 
\be
\widetilde{E}(\np) 
\equiv
f_0(\epsilon(\np),\np) 
= \sqrt{\np^2+\widetilde{m}^2(\np)}
\label{Dirac-energy}
\ee
represents the nucleon's Dirac energy.
The field strength
renormalization constant, $\sqrt{Z_2(\np)}$, in Eq.~(\ref{phispin}) 
is obtained from the renormalized nucleon propagator~\cite{Ama02,Pes95}
and reads
\begin{eqnarray}
Z_2(\np)&=& {\rm Res}
\left. \frac{1}{[1-C(P)][f_0(P)-\widetilde{E}(P)]}\right|_{p_0=\epsilon(\np)}
\nonumber\\
&=& \left[ 1-B
-p_0 \frac{\partial B}{\partial p_0}
-m\frac{\widetilde{m}}{\widetilde{E}}\frac{\partial A}{\partial p_0} 
+\frac{\np^2}{\widetilde{E}}\frac{\partial C}{\partial p_0} 
\right]_{p_0=\epsilon(\np)}^{-1}\ .
\label{Z2}
\end{eqnarray}

In the relativistic Hartree-Fock model the free spinors are used to
compute the first approximation to the self-energy. This is then
inserted in the Dirac equation to get new spinors, and so on.  This
self-consistent procedure has to be dealt with numerically.

The resulting momentum distribution is then obtained with the renormalized wave functions
\be
\widetilde\psi_{\nk,s}(\np) =  \int_V d\nr\, 
\widetilde\psi_{\nk,s}(\nr) e^{-i\np\cdot\nr} 
= \sqrt{\frac{V \widetilde m(\np)}{\widetilde E(\np)}}\, 
\widetilde u_s(\np,\widetilde m(\np)) \delta_{\nk,\np}
\label{psitilde}
\ee
and reads
\be
\widetilde n(\np) = \sum_{\nk,s}
\widetilde\psi_{\nk,s}^\dagger(\np)\widetilde\psi_{\nk,s}(\np)
\theta(\widetilde k_F
-p)\ ,
\label{nptilde}
\ee
where $\widetilde k_F$, $\widetilde
m(\np)$ and $\widetilde E(\np)$ are the nucleon's renormalized Fermi momentum,
mass and energy, respectively.
From Eqs.~(\ref{nptilde},\ref{psitilde},\ref{phispin}) 
the HF momentum distribution is then found to be
\be
\widetilde n(\np) = 2 V Z_2(\np) \theta(\widetilde k_F-p)\ ,
\label{npHF}
\ee
which clearly reproduces the free result in 
Eq.~(\ref{npfree}) for $Z_2(\np)=1$ and $\widetilde k_F=k_F$.

Note that the HF wave function with the spinor~(\ref{phispin}) is {\em not} normalized to
unity. Indeed 
\be
\int_V d\nr \widetilde \psi_{\nk,s}^\dagger(\nr) \widetilde\psi_{\nk,s}(\nr)
= \frac{\widetilde m(\np)}{\widetilde E}\widetilde
u_s^\dagger(\nk,\widetilde m(\np)) \widetilde u_s(\nk,\widetilde m(\np))
= Z_2(\nk)\ .
\label{norm}
\ee
However the total number of nucleons must be conserved.
This implies that the unitarity condition
\be  
\int \frac{d\np}{(2\pi)^3} \widetilde n(\np)
= 2 V \int \frac{d\np}{(2\pi)^3} Z_2(\np) \theta(\widetilde k_F-p) = Z
\label{intZ2}
\ee
must be fulfilled. Equation~(\ref{intZ2}) can be viewed as 
the procedure to fix the HF Fermi momentum
$\widetilde k_F$, which can in principle be different from the free one.

Here we consider the first-order correction to the
momentum distribution arising from the HF series.
We shall focus on mesonic correlations, induced by the exchange of
$\pi$, $\rho$, $\omega$ and $\sigma$, associated with the following interaction
Lagrangian~\cite{Mac87}
\be
{\cal L}(x) = \bar\psi(x) \left\{ 
\frac{g_\pi}{2 m} \gamma^5 \gamma^\mu \ntau \cdot \partial_\mu \npi(x)
+ g_\rho \left[\gamma^\mu - \frac{a_\rho}{m} \sigma^{\mu\nu}
\partial_\nu \right] \ntau \cdot \nrho_\mu (x)
+ g_\omega \gamma^\mu \omega_\mu(x)
+ g_\sigma \sigma(x)
\right\} \psi(x) \ .
\ee

Using this Lagrangian we compute the self-energy in OBE approximation.
For each meson $i=\pi,\rho,\omega,\sigma$,
the corresponding self-energy functions $A_i(P),B_i(P),C_i(P)$  
are given in the Appendix. 
The total self-energy is obtained from 
Eq.~(\ref{spin}) with $A(P)=\sum_i A_i(P)$, $B(P)=\sum_i B_i(P)$
and  $C(P)=\sum_i C_i(P)$.
While the pion and rho self-energies correspond to purely exchange (Fock) terms,
the sigma and omega also have a direct (Hartree) contribution due to
their isoscalar nature.

The HF energy $\epsilon(\np)$, the solution of
Eq.~(\ref{dispersion}), can be computed analytically
to first order in the squared meson-nucleon coupling 
constant $g_i^2$.  
For this purpose we note that the functions $A_i(P)$, $B_i(P)$ and 
$C_i(P)$ are of order $O(g_i^2)$.  
Hence the following expansion of the Dirac mass in Eq.~(\ref{m-tilde}) holds:
\begin{equation}\label{m-expansion-1}
\widetilde{m}(P)=
m\left[1+A(P)+C(P)\right]+O(g_i^4)\ .
\end{equation}
Inserting this into Eq.~(\ref{dispersion}) 
and expanding the righthand side to first order in $g_i^2$, we get the equation
\begin{eqnarray}
p_0
&\simeq& 
E_\np+\Delta E(P)\ ,
\label{p0}
\end{eqnarray}
where 
\begin{eqnarray}
\Delta E(P) = 
\frac{1}{E_\np}\left[m^2 A(P)+E_\np^2 B(P)-\np^2 C(P)\right]\ .
\label{energycorrection}
\end{eqnarray}
Next we insert the value of $p_0$ given by Eq.~(\ref{p0}) 
inside the functions $A(P)$, $B(P)$, $C(P)$ and
expand them around the on-shell value $p_0=E_\np$, 
neglecting terms of second order in $g_i^2$. 
We get
\begin{equation}
A(P)\simeq A(E_\np+\Delta E,\np) \simeq
A(E_\np,\np)\equiv A_{0}(\np)
\end{equation}
and likewise $B(P) \simeq  B_{0}(\np)$,  $C(P) \simeq  C_{0}(\np)$.
Inserting these on-shell values into Eq.~(\ref{energycorrection}) 
we finally obtain the HF energy to first order:
\begin{equation}
p_0 \simeq E_\np+\frac{1}{E_\np} 
\left[m^2 A_{0}(\np) + E_\np^2 B_{0}(\np) 
-\np^2 C_{0}(\np) \right] =\epsilon(\np)\ .
\label{HFE}
\end{equation}

We proceed now by expanding as well
the renormalized wave function, see Eqs.~(\ref{psitilde},\ref{phispin}). 
For this purpose we expand the Dirac mass in Eq.~(\ref{m-expansion-1}) 
around the on-shell energy
\begin{equation}
\widetilde{m}(\np)
\simeq
m\left[1+A_{0}(\np)+C_{0}(\np)\right]
\label{m-expansion}
\end{equation}
and likewise the Dirac energy $\widetilde{E}(\np)$ defined in 
Eq.~(\ref{Dirac-energy})
\begin{eqnarray}
\widetilde{E}(\np)
&=& \frac{1-B}{1-C}\epsilon(\np)
\simeq
E_\np+\frac{m^2}{E_\np} 
\left[A_{0}(\np) + C_{0}(\np) \right]
\ .
\label{E-expansion}
\end{eqnarray}
Moreover for the field-strength renormalization of Eq.~(\ref{Z2})
we obtain
\begin{equation}
Z_2(\np) \simeq 1+\alpha(\np)
\end{equation}
with
\begin{equation} \label{alpha}
\alpha(\np) \equiv
 B_0(\np)
+\left[\frac{m^2}{E_\np}\frac{\partial A}{\partial p_0} 
+E_\np\frac{\partial B}{\partial p_0}
-\frac{\np^2}{E_\np}\frac{\partial C}{\partial p_0} 
\right|_{p_0=E_\np}\ .
\end{equation}

After some algebra the following first-order expression for the
HF wave function is obtained
\begin{equation} \label{A1}
\widetilde \psi_{\nk,s}(\np)
\simeq
\sqrt{\frac{m}{E_\np}}
\left[ 1
     +m\frac{A_0(\np)+C_0(\np)}{E_\np}
      \frac{E_\np\gamma_0-m}{2E_\np}
     +\frac12\alpha(\np)
\right] u_s(\np,m) \delta_{\nk,\np}\ .
\end{equation}

The above expansion transparently displays the effect of the self-energy on 
the free wave function.
Indeed the second term in the square brackets of Eq.~(\ref{A1}) corresponds 
to a negative-energy component with momentum $\np$. 
Thus, within the OBE potential approach the renormalized HF spinors in the
nuclear medium are characterized by two new elements with respect to
the bare $u_s(\np,m)$: the term
$(E_\np\gamma_0-m)u_s(\np,m)$, directly connected with the negative-energy
components in the wave function, and the term $\alpha(\np)$, arising from
the field strength renormalization $\sqrt{Z_2(\np)}$.
However, the negative energy component does not contribute
to the momentum distribution in first order, where one simply gets
\be
\widetilde n(\np)\simeq 2 V \left[1+\alpha(\np)\right]\theta(k_F-p) \ .
\label{np1}
\ee

The explicit expression for the first order expansion of the 
function $\alpha(\np)\equiv\alpha(p)=\sum_i\alpha_i(p)$ is
given in the Appendix.
Note that the Hartree self-energy of the $\omega$ and $\sigma$ 
does not contribute to $\alpha$.
The unitarity condition of Eq.~(\ref{intZ2}) becomes
\ba
2 V \int \frac{d\np}{(2\pi)^3} Z_2(\np) \theta(\widetilde k_F-p) &=&
\frac{V\widetilde k_F^3}{3\pi^2}+\frac{V}{\pi^2} 
\int_0^{\widetilde k_F} p^2 dp \alpha(p) = Z\ ,
\label{intZ2pi}
\ea
which is certainly satisfied by $\widetilde k_F=k_F$, because
the function $\alpha$ exactly satisfies 
\begin{equation}
\label{unitarity}
\int_0^{k_F}{\rm d}p\, p^2 \alpha(p)=0
\end{equation}
(see Appendix). 
We have numerically checked that there is no other value of
$\widetilde k_F$ for which the number of particles is $Z$.
Therefore {\em to first order in $g_i^2$ the Fermi momentum is not affected 
by the Hartree-Fock field}. This means that the present calculation
respects not only Lorentz covariance, but also unitarity.

\begin{figure}[th]
\begin{center}
\leavevmode
\def\epsfsize#1#2{0.7#1}
\epsfbox[160 320 440 720]{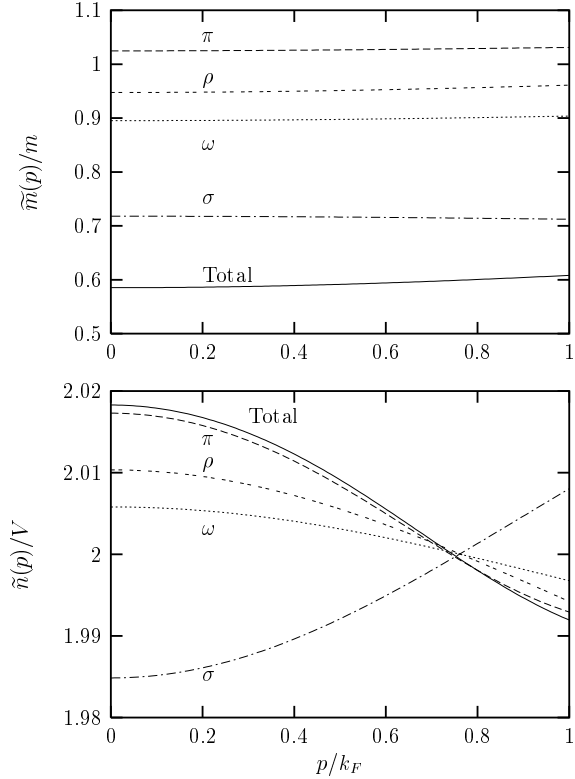}
\end{center}
\caption{The ratio $\widetilde m(\np)/m$ of Eq.~(\protect\ref{m-expansion}) 
(upper panel)
and the momentum distribution per unit volume of Eq.~(\protect\ref{np1}) 
(lower panel) are plotted
versus the nucleon momentum divided by the Fermi momentum ($k_F$=250 MeV/c). 
The results obtained by taking into account one single meson
(pion: dashed; rho: double-dashed, omega: dotted; sigma: dot-dashed)
are displayed together with the
total result (solid). The mesonic parameters are~\protect{\cite{Mac01}}: 
$m_\pi$=139.6 MeV/c,
$m_\rho$=770 MeV/c,
$m_\omega$=782 MeV/c,
$m_\sigma$=550 MeV/c,
$g_\pi^2/4\pi^2$=13.6,
$g_\rho^2/4\pi^2$=0.84, $a_\rho$=6.1,
$g_\omega^2/4\pi^2$=20,
$g_\sigma^2/4\pi^2$=7.78,
$\Lambda_\pi$=1720 MeV/c,
$\Lambda_\rho$=1310 MeV/c,
$\Lambda_\omega$=1500 MeV/c,
$\Lambda_\sigma$=2500 MeV/c .
}
\end{figure}

In Fig.~1 we plot the first order Dirac mass of Eq.~(\ref{m-expansion}) 
(top panel) and the first order momentum distribution of
Eq.~(\ref{np1}) (bottom panel) as functions of $p/k_F$ for $k_F$=250 MeV/c.
The separate contributions of the various mesons
are displayed. In the present calculation we empirically account for
the short-range physics through the meson-nucleon form factors 
$F_i(\nk)=\frac{\Lambda_i^2-m_i^2}{\Lambda_i^2+\nk^2}$ which cut off the
nucleon-nucleon interaction in a spatial region of size $\sim 1/\Lambda_i$.
Actually, for sake of simplicity, we have
approximated their effect by multiplying the self-energy 
associated with each meson by a constant factor (1 for the pion, 0.9 for the
sigma, 0.5 for the omega and 0.4 for the rho): the form factors are indeed 
slowly varying functions of the meson momentum in the integration domain.
The figure shows that the most sizable contribution to the
Dirac mass arises from the 
$\sigma$-meson, which reduces the mass by about 30$\%$, whereas the impact
of the other mesons is at most 10$\%$
(in particular the pion induces a negligible increase of the mass): the total
effect in the present model is a reduction of the mass by about a factor 0.6,
in accord with the findings of Refs.~\cite{Ser86,Ana81,Sch01}. It is also
remarkable that the $p$-dependence of the Dirac mass is almost negligible.

As far as the momentum distribution is concerned, it appears that the
$\sigma$, carrying an attractive interaction,
induces a depletion of the baryonic density at low
momenta and an enhacement of the latter in the vicinity of the Fermi
surface, in contrast with the effect of the other mesons.
It is interesting to note that the size of the
Fock contribution in the momentum distribution decreases as the meson
mass increases. This is in agreement with the fact that, at least for quasielastic inclusive electroweak responses modeled as we do here, the forces carried by the heavier mesons can be reasonably well approximated by four-fermion point interactions.
In this case the HF approximation can be expressed as a linear combination of
Hartree terms, which, as previously mentioned, do not affect the 
momentum distribution. Furthermore, and notably, the contributions arising 
from $\rho$, $\sigma$ and $\omega$ cancel almost exactly.
Thus the net effect of the full interaction
coincides with the one obtained with the pion alone, and it amounts
to an
increase of the nucleon momentum density by about 1\% for $p\simeq 0$ and to a
decrease of it by almost the same amount for $p\simeq k_F$, in such a
way that the number of nucleons is conserved, according to the
unitarity condition of Eq.~(\ref{intZ2}). 
It is worth noticing that the reduction of the
momentum distribution near the Fermi surface due to relativistic HF
correlations is of the same size as the one arising from 
short-range correlations of Jastrow type only,
found in Refs.~\cite{Fab01,Maz02} in the spectroscopic
factors of quasi-hole valence states.  Note, however, that this
effect is very small compared with that expected from 
a more sophysticated non-relativistic modeling of short-range 
correlations \cite{Mut00}, 
although one cannot make this statement with certainty in a relativistic context, since a relativistic version of Brueckner HF is even more challenging 
to carry out than relativistic HF and both constitute work for the future.

\begin{figure}[th]
\begin{center}
\leavevmode
\def\epsfsize#1#2{0.7#1}
\epsfbox[160 320 440 720]{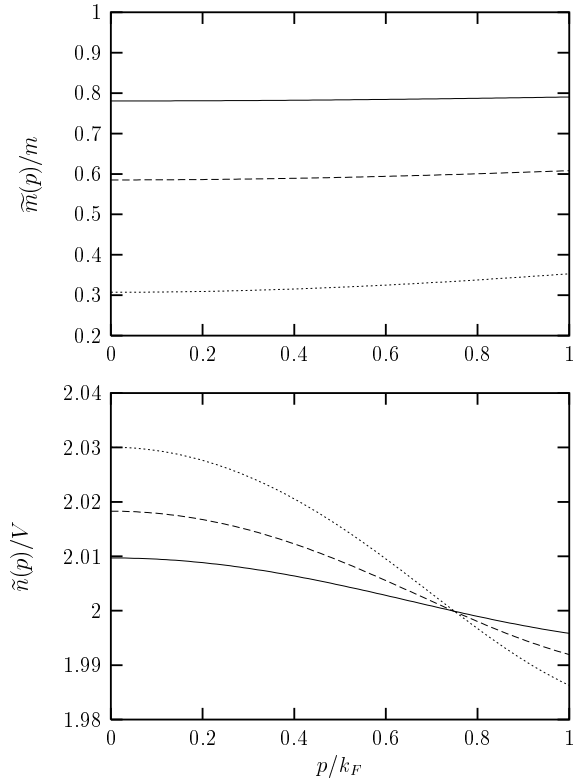}
\end{center}
\caption{The ratio $\widetilde m(\np)/m$ of Eq.~(\protect\ref{m-expansion}) 
(upper panel)
and the momentum distribution per unit volume of Eq.~(\protect\ref{np1}) 
(lower panel) are plotted
versus $p/k_F$ for $k_F$=200 MeV/c (solid), 
 $k_F$=250 MeV/c (dashed) and $k_F$=300 MeV/c (dotted). 
}
\end{figure}

In Fig.~2 the same observables displayed in Fig.~1, 
the contributions of all the mesons being included, are shown for three values of the Fermi
momentum: $k_F=200$ (solid), 250 (dashed) and 300 (dotted) MeV/c.
It appears that the effect of the mesonic HF field on both the Dirac
mass and the momentum distribution increases with the density and it is
roughly proportional to $k_F^3$.
However the origin of this dependence is different in the two cases, due to the
different role played by the various mesons.
Indeed the Dirac mass essentially stems from the $\sigma$ and $\omega$,
whereas the baryonic momentum density is significantly affected by all the four mesons.

A simple analysis of the $k_F$-dependence of the OBE contributions to
$\widetilde m(\np)$ and $\widetilde n(\np)$ can be performed
through an expansion in the small parameter $\eta_F=k_F/m$,
whose typical value is $\sim$ 1/4. 
Such an expansion has been successfully applied to the study of inclusive
and exclusive electron scattering both for free and correlated nuclear 
systems~\cite{Ama96,Jes98,Ama98,Ama99,Alv01}. 
Remarkably, the $\eta_F$ expansion has been shown to be very useful for 
exploring the role of chiral pion dynamics in nuclear matter~\cite{Kai02_1,Kai02_2}. 

When performing this expansion one should pay attention to the fact that the
pion is much lighter than the other mesons: this induces a different 
$k_F$-dependence for the pionic contributions, since $k_F/m_\pi$ cannot
be treated as a small parameter. In fact it is easy to show
that the heavy mesons' contributions to 
$\widetilde m(\np)$ go as $k_F^3$ (the
pion contribution is negligible). On the other hand in 
$\widetilde n$
the pionic effect grows as $k_F^3$, while the heavy
mesons contribute as $k_F^5$. We recall that the $\sigma$, $\omega$ and $\rho$ 
almost cancel in the momentum distribution (see Fig.~1).

It is of importance to notice that the physics of real nuclei roughly corresponds
to the range $200\leq k_F\leq 250$ MeV/c: here our prediction for the Dirac mass,
$\widetilde{m}$, is close to the one for the effective mass~\cite{Cel86,Mar96}.
It is only for larger $k_F$ that the two quantities start to differ substantially.

Finally, it is also interesting to note that the momentum distribution $\widetilde n(p)$
coincides with the free one, $n(p)=2$, for a value of $p/k_F$, 
which is independent of both the specific meson and the value of $k_F$ 
(see Figs.~1 and 2). This finding can
again be interpreted on the basis of the above-mentioned expansion,
which shows that the function $\alpha$ vanishes for $p\simeq \sqrt{3/5} k_F$.

Before drawing our conclusions we would like to address the issue
of the relevance of our findings on physical observables. In this regard
we have shown in Refs.~\cite{Ama01,Ama02} that the effect on the
electromagnetic response functions including pionic correlations
due to the modification of the momentum
distribution is negligible (see in particluar Fig.~13 in \cite{Ama02}).
One could not say this a priori and so an important conclusion of that work plus the deeper understanding presented in this letter is that at the level where relativity is dealt with in a consistent way such correlations appear under typical circumstances to be perturbatively small.

In summary, in this letter we have presented a relativistic analysis
of the single-particle properties of nuclear matter in HF
approximation within a meson-exchange model.  In particular, we have
focused on the role played by the pion, rho, omega and sigma on the
Dirac mass of the nucleon and on the momentum distribution.  Whereas
the momentum distribution is not affected by HF field in a
non-relativistic framework, in the relativistic case it is slightly
modified due to the renormalization of the spinors. In this work we
have quantified this effect to first order in the coupling constant
where the HF equations can be solved analytically.  Using this
solution we have demonstrated that the field strength renormalization
function exactly satisfies unitarity at this order.

Moreover, whereas for the  Dirac mass, as is well known, the effect
of HF mesonic correlations amounts to about 30-40\% and mainly arises
from the $\sigma$ and
$\omega$ mesons, we have shown that in the momentum distribution 
a cancellation among
the heavier mesons occurs and the total result basically coincides
with the pionic contribution, which amounts to a 1-3\% effect,
depending upon the density.

\subsection*{Acknowledgments}

This work was partially supported by funds provided by DGICYT (Spain)
under Contract Nos. PB/98-1111, PB/98-0676 and PB/98-1367 and the Junta de 
Andaluc\'{\i}a (Spain) and by the INFN-CICYT exchange, as well as by
the U.S. Department of Energy under cooperative research agreement
No. DE-FC02-94ER40818. M.B.B. acknowledges MEC (Spain)
for a sabbatical stay at University of Sevilla (Ref. SAB2001-0025).

\appendix

\section{Appendix}

The functions $A,B,C$ of Eq.~(\ref{spin})
can be expressed in terms of the integrals
\begin{eqnarray}
I(P,m_i) &\equiv&
\int\frac{d^3 k}{(2\pi)^3}\theta(k_F-k)
\frac{1}{2E_\nk}
\frac{1}{(P-K)^2-m_i^2} \ ,
\label{I}
\\
L^{\mu}(P,m_i) &\equiv&
\int\frac{d^3 k}{(2\pi)^3}\theta(k_F-k)
\frac{1}{2E_\nk}
\frac{K^{\mu}}{(P-K)^2-m_i^2}\ ,
\label{Lmu}
\\
L^{\mu\nu}(P,m_i) &\equiv&
\int\frac{d^3 k}{(2\pi)^3}\theta(k_F-k)
\frac{1}{2E_\nk}
\frac{K^{\mu} K^{\nu}}{(P-K)^2-m_i^2}\ ,
\label{Lmunu}
\end{eqnarray}
through the following relations
\footnote{
${\bf L}$ is parallel to ${\bf p}$ since, choosing ${\bf p}$ along the $z$
axis, the azimuthal integration in Eq.~(\ref{Lmu}) yields $L_1=L_2=0$.}:
\begin{eqnarray}
A_\pi(P) &=&
\frac{3 g_\pi^2}{2}
\left[I(P,m_\pi)-\frac{P_\mu L^\mu(P,m_\pi)}{m^2}+
\frac{P^2-m^2}{2m^2} I(P,m_\pi)
\right] \ ,
\label{Api-definition}\\ 
B_\pi(P) &=&
\frac{3 g_\pi^2}{2}
\left[I(P,m_\pi)-\frac{P_\mu L^\mu(P,m_\pi)}{m^2}+
\frac{P^2-m^2}{2m^2}
\frac{L_0(P,m_\pi)}{p_0}
\right] \ ,
\label{Bpi-definition}\\ 
C_\pi (P) &=&
\frac{3 g_\pi^2}{2}
\left[I(P,m_\pi)-\frac{P_\mu L^\mu(P,m_\pi)}{m^2}+
\frac{P^2-m^2}{2m^2}\frac{L_3(P,m_\pi)}{p}
\right] 
\label{Cpi-definition}
\end{eqnarray}
for the pion,
\begin{eqnarray}
A_\rho(P) &=&
6 g_\rho^2
\left[(2+3 a_\rho+3a_\rho^2)I(P,m_\rho)
-3a_\rho(1+a_\rho)\frac{P_\mu L^\mu(P,m_\rho)}{m^2}
\right.
\nonumber\\
&+&
\left.\frac{3a_\rho^2}{2m^2}(P^2-m^2)I(P,m_\rho)
\right] \ ,
\label{Arho-definition}\\ 
B_\rho(P) &=&
6 g_\rho^2
\left\{(3 a_\rho+2a_\rho^2)I(P,m_\rho)-
\frac{a_\rho^2}{m^2} P_\mu 
\left[2 L^\mu(P,m_\rho)-\frac{L^{0\mu}(P,m_\rho)}{p_0}\right]\right.
\nonumber\\
&-&
\left.
\left[1+3a_\rho+a_\rho^2-\frac{a_\rho^2}{2m^2}(P^2-m^2)\right]
\frac{L_0(P,m_\rho)}{p_0}\right\} \ ,
\label{Brho-definition}\\ 
C_\rho(P) &=& 6 g_\rho^2 \left\{(3 a_\rho+2a_\rho^2)I(P,m_\rho)-
\frac{a_\rho^2}{m^2} P_\mu \left[2
L^\mu(P,m_\rho)-\frac{L^{3\mu}(P,m_\rho)}{p} \right] \right.
\nonumber\\ &-& \left.
\left[1+3a_\rho+a_\rho^2-\frac{a_\rho^2}{2m^2}(P^2-m^2)\right]
\frac{L_3(P,m_\rho)}{p}\right\}
\label{Crho-definition}
\end{eqnarray}
for the rho,
\begin{eqnarray}
A_\omega(P) &=&
4 g_\omega^2 I(P,m_\omega) \ ,
\label{Aomega-definition}\\ 
B_\omega(P) &=&
2 g_\omega^2
\left[\frac{k_F^3}{3 p_0\pi^2
m_\omega^2}-\frac{L_0(P,m_\omega)}{p_0}\right]\ ,
\label{Bomega-definition}\\ 
C_\omega(P) &=&
-2 g_\omega^2
\frac{L_3(P,m_\omega)}{p}
\label{Comega-definition}
\end{eqnarray}
for the omega and
\begin{eqnarray}
A_\sigma(P) &=&
- g_\sigma^2 \left[I(P,m_\sigma)+\frac{1}{\pi^2 m_\sigma^2}
\left(k_F E_F-m^2\ln\frac{k_F+E_F}{m}\right)\right] \ ,
\label{Asigma-definition}\\ 
B_\sigma(P) &=&
- g_\sigma^2 \frac{L_0(P,m_\sigma)}{p_0} \ ,
\label{Bsigma-definition}\\ 
C_\sigma(P) &=&
- g_\sigma^2
\frac{L_3(P,m_\sigma)}{p}
\label{Csigma-definition}
\end{eqnarray}
for the sigma.

The corresponding expression for the functions $\alpha_i$
(see Eq.~(\ref{alpha})), with
$\alpha=\sum_{i=\pi,\rho,\omega,\sigma} \alpha_i$, is
\begin{equation}
\label{alphai}
\alpha_i(p)=
\frac{m_i^2 g_i^2}{4\pi^2 E_\np} 
\int_0^{k_F} dk \frac{k^2}{E_\nk}
\,\frac{E_\nk-E_\np}{\gamma_i^2(p,k)-4 p^2 k^2}
\, f_i(p,k) \ ,
\end{equation}
where
\begin{equation}
\gamma_i(p,k) \equiv (E_\np-E_\nk)^2-
p^2-k^2-m_i^2=2m^2-m_i^2-2E_\np E_\nk
\label{gamma}
\end{equation}
and  the functions $f_i(p,k)$ are defined as 
\begin{eqnarray}
\label{alphapi}
f_\pi(p,k) &=& 3 \ ,
\\
f_\rho(p,k) &=& 3
\left[
2(1+6 a_\rho+4 a_\rho^2)+a_\rho^2\frac{m_\rho^2}{m^2}+4\frac{m^2}{m_\rho^2}
\right.
\nonumber
\\
&+& \left. 
\left(2+6a_\rho+a_\rho^2\frac{m_\rho^2}{m^2}\right)
\frac{\gamma_\rho^2(p,k)-4 p^2 k^2}{4 k p m_\rho^2 }
\ln\frac{\gamma_\rho(p,k)+2 k p}{\gamma_\rho(p,k)-2 k p}\right]
\ ,
\\
f_\omega(p,k) &=& 2
\left[1+2\frac{m^2}{m_\omega^2}
+ \frac{\gamma_\omega^2(p,k)-4 p^2 k^2}{2 k p m_\omega^2} 
\ln\frac{\gamma_\omega(p,k)+2 k p}{\gamma_\omega(p,k)-2 k p}\right]
\ ,
\\
f_\sigma(p,k) &=& 
1-4\frac{m^2}{m_\sigma^2}
+\frac{\gamma_\sigma^2(p,k)-4 p^2 k^2}{4 k p m_\sigma^2} 
\ln\frac{\gamma_\sigma(p,k)+2 k p}{\gamma_\sigma(p,k)-2 k p}
\ .
\end{eqnarray}
The meson-nucleon form factors have been neglected for simplicity. Their
impact on the results is discussed in the text.  Using the above
expressions, the unitarity condition of Eq.~(\ref{unitarity}) follows
since the functions $f_i(p,k)$ are all symmetrical under the exchange of
$p$ and $k$, hence
\begin{equation}
\int_0^{k_F} dp p^2 \alpha_i(p) 
= 
\frac{m_i^2 g_i^2}{4\pi^2}
\int_0^{k_F} dp 
\int_0^{k_F} dk \,\frac{p^2k^2}{ E_\np E_\nk}
\,\frac{E_\nk-E_\np}{\gamma_i^2(p,k)-4 p^2 k^2}
\, f_i(p,k) 
=0 \ .
\end{equation}



\begin{thebibliography}{Expo92}

\bibitem{Fet71}  A.L. Fetter, J.D. Walecka,
                 Quantum Theory of Many-Particle Systems,
                 McGraw-Hill, New York, 1971.

\bibitem{Mut00}  H. M\"uther, A. Polls, 
                 Prog. Part. Nucl. Phys. {\bf 45} (2000) 243.

\bibitem{Fan84}  S. Fantoni, V. Pandharipande,
                Nucl. Phys. A 427 (1984) 473.

\bibitem{Dic92} W. Dickhoff, H. M\"uther, 
                Rep. Prog. Phys. {\bf 55} (1992) 1947.
 
\bibitem{Ama01} J.E. Amaro, M.B. Barbaro, J.A. Caballero, 
                T.W. Donnelly, A. Molinari,
                Nucl. Phys. {\bf A 697} (2002) 388.

\bibitem{Ama02} J.E. Amaro, M.B. Barbaro, J.A. Caballero, 
                T.W. Donnelly, A. Molinari,
                Phys. Rep. {\bf 368} (2002) 317.

\bibitem{Ser86} B.D. Serot, J.D. Walecka,
                Adv. Nucl. Phys. {\bf 16} (1988) 1.

\bibitem{Cel86}  L.S. Celenza, C. M. Shakin,
                 Relativistic Nuclear Physics,
                 World Scientific, Singapore, 1986.

\bibitem{Bjo65}  J.D. Bjorken, S.D. Drell,
                 Relativistic Quantum Mechanics,
                 McGraw-Hill, New York, 1965.

\bibitem{Ana81}  M.R. Anastasio, L.S. Celenza, C.M. Shakin,
                 Phys. Rev. {\bf 23} (1981) 569.

\bibitem{Pes95}  M.E. Peskin, D. V. Schroeder,
                 An introduction to quantum field theory,
                 Perseus, 1995.

\bibitem{Mac87} R. Machleidt, K. Holinde, Ch. Elster,
                Phys. Rep. {\bf 149}, No. 1 (1987) 1.

\bibitem{Sch01} E. Schiller, H. M\"uther,
                Eur. Phys. Jour. {\bf A11} (2001) 15.

\bibitem{Fab01} A. Fabrocini, G. Co',
                Phys. Rev. {\bf C63}, 044302 (2001). 

\bibitem{Maz02} M. Mazziotta, J.E. Amaro, F. Arias de Saavedra, 
                Phys. Rev. {\bf C65}, 034602 (2002). 

\bibitem{Ama96} J.E. Amaro, J.A. Caballero, T.W. Donnelly,
                A.M. Lallena, E. Moya de Guerra, J.M. Udias,
                Nucl. Phys. {\bf A602} (1996) 263.

\bibitem{Jes98} S. Jeschonnek, T.W. Donnelly, 
                Phys. Rev. {\bf C57} (1998) 2438.

\bibitem{Ama98} J.E. Amaro, M.B. Barbaro, J.A. Caballero, 
                T.W. Donnelly, A. Molinari,
                Nucl. Phys. {\bf A643} (1998) 349.

\bibitem{Ama99} J.E. Amaro, M.B. Barbaro, J.A. Caballero, 
                T.W. Donnelly, A. Molinari,
                Nucl. Phys. {\bf A657} (1999) 161.

\bibitem{Alv01} L. \'Alvarez-Ruso, M.B. Barbaro, T.W. Donnelly, A. Molinari,
                Phys. Lett. {\bf B497} (2001) 214.

\bibitem{Kai02_1} N. Kaiser, S. Fritsch, W. Weise, 
                 Nucl.Phys. {\bf A700} (2002) 343.

\bibitem{Kai02_2} N. Kaiser, S. Fritsch, W. Weise,
                 Nucl.Phys. {\bf A697} (2002) 255.

\bibitem{Mar96} M.B. Barbaro, A. De Pace, T.W. Donnelly, A. Molinari,
                Nucl. Phys. {\bf A596} (1996) 553.

\bibitem{Mac01} R. Machleidt,
                Phys. Rev. {\bf C63}, 024001 (2001). 

\end{thebibliography}
\end{document}